# First principles modeling of oxygen adsorption on LaMnO$_3$ (001) surface

Eugene A. Kotomin[a,b*], Yuri A. Mastrikov[a,b], Eugene Heifets[a] and Joachim Maier[a]

**We present and discuss the results of *ab initio* DFT plane-wave supercell calculations of the atomic and molecular oxygen adsorption and diffusion on the LaMnO$_3$ (001) surface which serves as a model material for a cathode of solid oxide fuel cells. The dissociative adsorption of O$_2$ molecules from the gas phase is energetically favorable on surface Mn ions even on a defect-free surface. The surface migration energy for adsorbed O ions is found to be quite high, 1.6 eV. We predict that the adsorbed O atoms could penetrate into electrode first plane when much more mobile surface oxygen vacancies (migration energy of 0.69 eV) approach the O ions strongly bound to the surface Mn ions. *Ab initio* thermodynamics predicts that at typical SOFC operation temperatures (~1200 K) the MnO$_2$ (001) surface with adsorbed O atoms is the most stable in a very wide range of oxygen gas pressures (above 10$^{-2}$ atm).**

### Introduction

Optimization of materials for cathodes of solid oxide fuel cells (SOFC) is a scientifically challenging and technologically important problem [1]. A necessary prerequisite of systematic research is understanding of the mechanism of oxygen reduction and in particular, the identification of the rate determining steps [2]. Despite considerable experimental efforts, many fundamental questions remain open. It is in particular the surface diffusion step which is poorly understood and which is hard to tackle experimentally. At the moment, a most popular SOFC cathode materials is Sr-doped LaMnO$_3$ (LSM). This material attracted recently considerable attention also due to its applications in spintronics and magnetic cooling [3] which stimulated several theoretical studies of magnetic properties of pure LSM surfaces [4]. Recently we performed a series of studies of the (100) and (110) LaMnO$_3$ (LMO) surfaces focused on atomic, electronic and magnetic structures ([5] and references therein).

However, we are aware only of two literature reports that attempted to model at the *ab initio* level an interaction of oxygen with LMO and LSM surfaces [6]. In these calculations, a strongly polar (110) surface was chosen which consists of alternating planes LaMnO/O$_2$/LaMnO/… with opposite charges +4e, -4e, +4e…(per unit cell area). Moreover, the surface unit cell was very small and coincides with the bulk unit cell. This is why performed modelling of the O$_2$ molecule on the LaMnO-terminated surface in fact has very little in common with a molecular adsorption but much more with a growth of the proper oxygen surface plane above LaMnO plane terminating the slab used. As is well known [7], stabilization of such polar surfaces could be achieved by 50 % reduction of the first plane charge due to self-consistent charge redistribution. That is, the O$_2$-terminated surface is expected to have instead of -4e a charge of -2e (per unit cell), what was exactly observed by the authors [6] for the dissociation products of a O$_2$ molecule.

In this Communication, we model O adsorption on a much less polar LMO (001) surface consisting of alternating MnO$_2$/LaO/… planes with nominal charges ± 1e, and use a large surface unit cell corresponding to 12.5 % of surface coverage by an adsorbate which corresponds to the SOFC operation conditions [1]. Oxygen reduction is complex multi-step process. We focus here only on the first stage of the O interaction with cathode surface which is known to be rate-determining step.

We found optimal sites for atomic and molecular adsorption, O$_2$ dissociation, and O atom surface diffusion. The final step of oxygen interaction with the SOFC cathode is the penetration of the adsorbed O atoms into the electrode when they encounter surface oxygen vacancies. This is why we calculated also the migration energy for surface and bulk oxygen vacancies. Since the SOFC operates at high temperatures, we complemented our DFT calculations by the thermodynamic analysis of the LMO surface stability at such temperatures. Due to electrical neutrality, motion of charged surface defect is accompanied by the motion of surface electronic charge. As the latter does not significantly contributes to the activation energy, our results are transformable to SOFC fuel cell conditions in which in the steady state electrodes are supplied from the outer circuit.

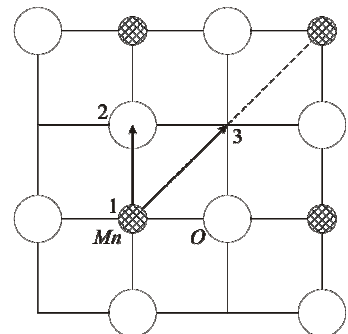

**Fig.1**. Schematic top view of three possible oxygen adsorption sites on MnO$_2$ (001) surfaces: over Mn (1), O (2) and the hollow position (3). If O atom is adsorbed atop surface Mn, the possible saddle points of its migration could be positions 2 or 3.

### Method

We employed ab *initio* DFT plane-wave-based computer code VASP [8] with a plane wave basis set, PAW method to represent the atom cores and with the GGA-Perdew-Wang-91 exchange-correlation potential, employing the cut-off energy of 400 eV and 2×2×2 Monkhorst-Pack k-point mesh in the Brillouin zone [9]. Preliminary spin polarized calculations of the LMO bulk and the (001), (110) surfaces are in a good agreement with experiment (when available) [5]. In particular, for the experimental orthorhombic geometry the energetically most favourable is A-type anti-ferromagnetic (AFM) configuration, in agreement with experiment whereas the cubic lattice constant (stable above 750 K) exceeds only by 0.5 % the experimental one. The calculated cohesive energy of 30.7 eV is close to the experimental value (31 eV). Calculations of the surface relaxation and surface energies show a weak dependence on the magnetic configuration [5]. Keeping in mind that relevant effects (~0.1 eV [5,6]) are much smaller than the adsorption and migration energies under study (several eV) and that under operation SOFC conditions LMO is paramagnetic ($T_N$=140 K), we performed all calculations for the ferromagnetic (FM) state with collinear spins. Moreover, the FM slab has the lowest energy. Neglect of spin polarisation results in considerably errors in material properties. We found in particular that 7- and 8- plane slabs are thick enough and show main property convergence.

For modelling oxygen adsorption based on our studies of pure surfaces [9], we used here a 7-plane slab terminated on both sides with $MnO_2$ plane covered by the adsorbate. Use of such symmetrical slab permits to avoid a problem of the surface dipole moment [7]. The periodically repeated slabs were separated by a vacuum gap of 15.8 Å (which corresponds to the thickness of 8-plane slab). The chosen surface unit cell was the extended $2\sqrt{2} \times 2\sqrt{2}$ primitive unit cell, i.e. the surface coverage was 1/8 (12.5%).

Since our calculations correspond to 0 K, we used the relevant low-temperature *orthorhombic* unit cell with on-plane lattice constants $a$= 5.56 Å, $b$=5.61 Å optimized for the bulk. All atomic coordinates were allowed to relax. Based on our experience for O adsorption on the $SrTiO_3$ (001) surface [10], we studied oxygen adsorption over surface Mn, O ions and the hollow position (Fig.1). The topological (Bader) effective atomic charges were calculated according to method described in Ref. [11]. The oxygen atom adsorption energies $E_{ads}^{(at)}$ were calculated with respect to free oxygen atoms:

$$E_{ads}^{(at)}(O) = - \frac{1}{2} [E_{slab}^{(ads)}(O) - E_{slab} - 2 E^{(O)}], \quad (1)$$

and with respect to free oxygen molecule:

$$E_{ads}^{(m)}(O) = - \frac{1}{2} [E_{slab}^{(ads)}(O) - E_{slab} - E^{(O2)}], \quad (2)$$

where $E_{slab}^{(ads)}$ is the total energy of a fully relaxed slab with two-sided adsorbate (O or $O_2$), $E_{slab}$ is the same for a pure slab, $E^{(O)}$ is the energy of isolated oxygen atom in the ground triplet state, and $E^{(O2)}$ is the total energy of isolated oxygen molecule in the triplet state. The prefactors ½ before brackets and 2 before $E^{(O)}$ appear since the interface is modeled by a substrate slab with two equivalent surfaces and both $O_{ads}$ atoms and $(O_2)_{ads}$ molecule symmetrically positioned on both sides of the slab. The molecular adsorption energy was calculated in a similar way. The difference of $E_{ads}^{(at)}(O)$ and $E_{ads}^{(m)}(O)$ equals to the $O_2$ molecule binding energy.

**Results and Discussion**
**Adsorption positions and energies**

Table 1 shows a strong preference for O atom adsorption over the surface Mn ion, unlike the bridge position between Ti and O ions found for the isostructural $SrTiO_3$ [10]. The difference is in line with the oxidizability of $Mn^{3+}$ (compared to $Ti^{4+}$). Note that the top of the valence band in $LaMnO_3$ is largely due to Mn orbitals whereas the O orbitals generate the valence band top in $SrTiO_3$. The electron charge of 0.62 e is transferred to the adsorbed O atom from nearest surface ions (0.18 e from nearest Mn, 0.16 e from four nearest O ions and the rest 0.28 e from next-nearest ions).

The analysis of the electron density redistribution confirms that the O adsorption induces a quite local perturbation. In the total density map one can see very well the "wavy" atomic structure of the orthorhombic slab. As a result of O adsorption, the spin momentum of Mn ion is strongly reduced. The test calculations performed for the high-temperature *cubic* phase (T> 750 K) give qualitatively similar results. In particular, the adsorption energy atop the Mn ion is 4.14 eV, ~ 3 % larger than in the orthorhombic phase.

In order to check how use of the nonstoichiometric slab affects the adsorption energy, we repeated calculations for the stoichiometric cubic 8-plane slab. The O adsorption energy atop Mn ion increased by 0.25 eV, or 6 %. The adsorbed O charge practically does not change (-0.64 e and -0.61 e, respectively) whereas the Mn ion charges exceed those on the perfect surface by 0.21 e and 0.1 e for 7- and 8-plane slabs, respectively. The atomic charges on 7-plane cubic and orthorhombic surfaces are very close.

The adsorption position near the surface O ion is practically identical to the *bridge position* found to be energetically most favourable for the $SrTiO_3$ (001) surface [10]. For $LaMnO_3$, this configuration turns out to be energetically less favourable, and the position above the hollow point the most unfavourable. In these two cases the adsorbed O atom receives 0.3-0.5 e from the nearest surface ions. Keeping in mind that the effective charges in LMO bulk and on the surface are considerably reduced due to the covalent component in the Mn-O chemical bonding as compared to the nominal charges ([5] and heading in Table 1), the configuration above the O ion could be treated as formation of a kind of $O_2^{(2-)}$ peroxo-molecule [6] tilted by $50^0$ towards nearest Mn ion.

**Table 1.** Calculated adsorption properties for O atoms on $MnO_2$ (001) orthorhombic surface. Energies in eV, distances from nearest ions, in Å, spins in $\mu_B$. S, T stand for singlet and triplet. The effective atomic charges on a pure surface: 1.67 e (Mn), -1.17 e (O). 2x, 4x indicate a number of equivalent atoms.

| Site | $E_{ads}^{(at)}(O)$ | $E_{ads}^{(m)}(O)$ | Distance from $O_{ads}$ | | Charges | | | Spin | |
|---|---|---|---|---|---|---|---|---|---|
| | | | $O_s$ | $Mn_s$ | $O_s$ | $Mn_s$ | $O_{ads}$ | Mn | O |
| Mn | 4.02 | 1.07 | 2.55(4x) | 1.63 | -1.13 | 1.85 | -0.62 | 2.20 | S |
| O | 2.41 | -0.54 | 1.50[a] (2x) | 1.87 | -0.71 | 1.65 | -0.48 | 3.61 | S |
| Hollow | 0.59 | -2.36 | 3.28(2x) 3.18(2x) | --- --- | -1.16 (4x) | ---- | -0.32 | --- | T |

[a]The O-O dumbbell has an angle of 50° with the normal to the surface

The calculated adsorption energies with respect to an O atom in a free molecule, $E^m_{ads}$, Eq. (2), are collected in Table 1. The positive value is obtained only for O atom atop the Mn ion, where the energy gain due to adsorption of two O atoms is larger than the molecule dissociation energy. (Despite the fact that our calculations overestimate the dissociation energy of a $O_2$ molecule -- 5.9 eV vs experimental 5.12 eV [12] – this does not affect our conclusion.)

**Oxygen migration**

The most stable position of the adsorbed O atom is atop of the surface Mn ion. Based on our results, the adsorbed O atom migration could occur with the barrier located nearby the O surface ion and with the activation energy exceeding 1.6 eV. Since the adsorbed oxygen atoms turned out to be strongly bound to the surface Mn ions and thus are quite immobile, penetration of these O atoms into the first plane of fuel cell cathode can occur predominantly upon their encounter with the mobile surface oxygen vacancies. Following our $SrTiO_3$ study [14], we calculated the equilibrium and saddle points for oxygen vacancies in the bulk and on the $MnO_2$ terminated surface (see also [13]). In the latter case, two pairs of nearest Mn and La ions are strongly displaced from the vacancy (ca. 0.2 Å) whereas the two O ions towards the vacancy (0.32 Å). The negative charge of the missing O ions is spread over nearest ions, mostly Mn. The calculated vacancy migration energy is 0.67 eV, smaller than that calculated for the bulk (0.95 eV). The latter value is typical for $ABO_3$ perovskites (see e.g. experimental data [14]) whereas the reduced energy on the surface also is in line with the trend in our calculations for vacancies in $SrTiO_3$ [15]. More detailed results will be published elsewhere. The key point here is that the vacancy mobility is much higher than that of the adsorbed O atoms and thus vacancy migration along cathode surface enables fast O transport to the electrode.

**Molecular adsorption**

Along with the O atom adsorption, we calculated also the $O_2$ *molecular* adsorption over the energetically most favorable position atop the Mn ion in the two configurations: perpendicular and parallel to the surface (called hereafter *tilted* and *horizontal* in Table 2). The molecule binding energy is larger for the tilted adsorption (Table 2). The total charge of the adsorbed molecule is

**Table 2.** Calculated adsorption properties for $O_2$ molecules on $MnO_2$ surface. Notations as in Table 1.

| Oriented | $E_{ads}^{(m)}(O_2)$ | Distances | | Charges[b] | | | Spin | |
|---|---|---|---|---|---|---|---|---|
| | | O-O bond | O-$Mn_s$ | O(1) | O(2) | $Mn_s$ | $Mn$[c] | $O_2$ |
| Tilted | 1.13 | 1.36 | 1.86[a] | -0.29[a] | -0.13 | 1.78 | 3.12 | T |
| horizontal | 0.89 | 1.42 | 1.85 1.90 | -0.30 | -0.30 | 1.77 | 3.05 | D |

a)For O atom nearest to the surface, b) Atoms in $O_2$ molecule c)3.80 $\mu_B$ on a pure surface

-0.42 e and the bond length 1.36 Å. (The bond length 1.3 Å of a free $O_2$ molecule calculated for the cutoff energy and other parameters used in this study are slightly larger than experimental value of 1.21 Å [12]). The adsorbed molecule could be considered as a kind of the superoxo-radical $O_2^-$ [6].

In the horizontal configuration the total charge of the molecule is larger, -0.6 e, the bond length increases up to 1.42 Å and it is closer to the peroxo-radical $O_2^{2-}$. In both cases we observe chemisorption (unlike a weak physical adsorption of $O_2$ on $SrTiO_3$ (001) surface). A comparison of atomic and molecular adsorption energies (Tables 1 and 2), indicates that $2E_{ads}^{(m)}(O) > E_{ads}^{(m)}(O_2)$ for the most favorable adsorption site atop the Mn ion. This means that the *dissociative* molecular adsorption is favorable even on the defectless surface – in contrast to $SrTiO_3$ [10].

**Surface stability under SOFC operation conditions**

In order to analyze the relative stability of different $LaMnO_3$ surfaces under realistic operation conditions of the fuel cell, we performed the *ab initio* thermodynamic treatment, similar to that applied earlier to $BaZrO_3$ [16] and $SrTiO_3$ [17] surfaces. This method was developed following a general *ab initio* thermodynamic approach ([18-23] and references therein). The most stable surface at any considered oxygen and manganese chemical potentials has the lowest surface Gibbs free energy.

(Note that we do not consider vibrational entropy changes which are supposed to be small [18].) We considered LaO- and $MnO_2$-terminated (001) surfaces built of *cubic* (high temperature) unit cells; LaMnO-, $O_2$-, and O-terminated (110) surfaces, and, finally, $MnO_2$-terminated (001) surface with adsorbed O atom (hereafter $MnO_2$+O– terminated surface). Since we assume the equilibrium of the surface with the bulk where the chemical potentials of three crystal constituents (La, Mn, and O) are interrelated by the condition $\mu(LaMnO_3)=\mu(La)+\mu(Mn)+3\mu(O)$, only chemical potentials of two of these components are independent variables. Because oxygen atoms are in equilibrium in the surface and in the $O_2$ gas above the cathode surface and we have to account for a strong dependence of O chemical potential on the $O_2$ gas partial pressure and temperature, it is suitable to choose the O chemical potential as one of independent variables in the surface Gibbs free energy whereas the Mn chemical potential can be taken as another independent variable. Following [16,17], we calculate these quantities with respect to the energy per atom in Mn metal ($\Delta\mu(Mn)$) and O atom in a free $O_2$ molecule ($\Delta\mu(O)$).

**Table 3** Calculated formation enthalpies $\Delta E_f$ (eV) for bulk oxides and $LaMnO_3$ perovskite. The second column contains experimental formation enthalpies for the oxides and $LaMnO_3$ [24] at $T_0$ =298.15 K, $p_0$ =1 atm. The last column provides a correcting shift of the experimental Gibbs free energies for $O_2$ gas calculated using different oxides (Ref. [16,17]).

| Crystal | Calc. $E_f$, eV | Expt $\Delta H_f$, eV |
|---|---|---|
| $La_2O_3$ | -18.86 | -18.60 |
| $Mn_2O_3$ | -11.57 | -9.92 |
| MnO | -4.19 | -3.99 |
| $MnO_2$ | -7.03 | -5.39 |
| $Mn_3O_4$ | -16.70 | -14.37 |
| $LaMnO_3$ | -15.66 | -14.77 |
| $LaMnO_3$ (from $La_2O_3$ and $Mn_2O_3$ oxides) | -0.498 | -0.514 |

We plotted the phase diagram on the l.h.s. of Fig. 2 which shows stability regions of different surfaces. The colour area here is limited at the bottom by the line, where the chemical potential of La atoms in LMO becomes larger than in a metal, what corresponds to a La precipitation. On other sides, the coloured area is limited by lines, where the lowest surface Gibbs free energies become negative and crystal spontaneously disintegrates. Only four different surfaces from six considered could be stable: LaO-terminated (001) surface, $MnO_2$+O-.

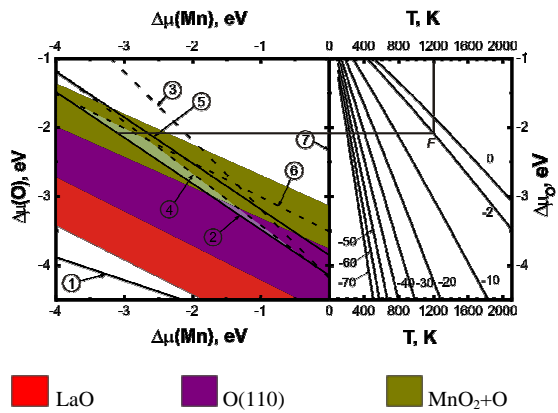

**Fig. 2** (Color). The phase diagram: the regions of stability of LaMnO$_3$ surfaces with different terminations as functions of chemical potential variation for Mn and O atoms. Comparison includes LaMnO-, O$_2$-, and O-terminated (011) surfaces, LaO- and MnO$_2$- terminated (001) surfaces, and MnO$_2$-terminated (001) surface with adsorbed O atom. The numbers in circles points to the lines, where metals or their oxides begin to precipitate: (1) metal La, (2) La$_2$O$_3$, (3) MnO$_2$, (4) Mn$_3$O$_4$, (5) Mn$_2$O$_3$, (6) MnO, and (7) metal Mn. The right side of the figures contains a family of $\Delta\mu_O$ as functions of the temperature at different oxygen gas pressures. The labels m on the oxygen lines indicate the gas pressure: $10^{-m}$ atm.

terminated (001) surface, O- and O$_2$- terminated (110) surfaces (the latter one is not seen in Fig. 4 within shown range of chemical potentials)

To determine the boundaries of the region, where precipitation of La- and Mn oxides does not occur, we computed the *formation energies* (enthalpies) for related metal oxides. Results for the oxide formation energies are provided in Table 3, where calculations are also compared with experimental formation enthalpies. Our computations slightly overestimate the formation energies, what is commonly observed for the DFT with GGA functionals. Still, when we considered formation of LaMnO$_3$ perovskite from the oxides with the same oxidation states as in the perovskites, the obtained value matches very well to that derived from experimental data.

The boundaries of the region, where LaMnO$_3$ crystal is stable with respect to decomposition into metals and their oxides, are represented in Fig.2 by lines 1 to 7. Pure LaMnO$_3$ surfaces could exist only between these lines [16,17]. The stability region is limited by line 2 (precipitation of La$_2$O$_3$) on the bottom and by lines 4 and 6 (precipitation of Mn$_3$O$_4$ and MnO, respectively) on the top. Because of the DFT shortcoming in calculations of the relative energies for materials with different degree of metal oxidation, one has to treat the obtained data with some caution. Therefore, we highlighted lines of precipitation for 3-valent metal oxides La$_2$O$_3$ and Mn$_2$O$_3$ (solid lines 2 and 5), where the oxidation state is the same as in LaMnO$_3$. From all considered surfaces, only *two* turn out to be stable (within the region of LaMnO$_3$ crystal stability): (i) MnO$_2$-terminated (001) surface with adsorbed O atoms (MnO$_2$+O surface) at high O chemical potentials (O-rich limit) and low Mn chemical potential (Mn-poor limit) and (ii) O-terminated (110) surface at low O chemical potentials (O-poor limit) and high Mn chemical potential (Mn-reach limit). This result remains valid, whatever we include into consideration only precipitation of 3-valent La and Mn oxides or also account for possible precipitation of all other manganese oxides.

On the r.h.s. of Fig. 2 we plotted the chemical potentials of the O atom derived from the experimental data (Ref. [24]). This window shows the dependence of the O chemical potential on temperature and O$_2$ partial pressure. Correct matching of the experimental curves with the computed stability diagram requires a correcting shift of the O atom chemical potential as explained in detail in ref. [16,17]. We calculated the shift employing the same set of oxides used in this work and thus obtained the average shift of -0.77 eV value used in Fig. 2.

The combination of our surface stability diagram with the oxygen chemical potential as a function of the temperature and O$_2$ partial pressure allows a detailed analysis of the trends under variable environmental conditions. Let us consider a situation typical for the fuel cell operation: T≈1200 K and oxygen partial pressure ranging between $p_{O2}$≈0.2 and 0.01atm (point *F* in Fig. 2). Our stability diagram demonstrates clearly that only the MnO$_2$-terminated (001) cubic surface with adsorbed O atoms is stable under the operational conditions (large O chemical potential, relatively high O pressure). The alternative (110) surface becomes dominant only at low O chemical potentials and thus very low oxygen pressures, $p_{O2} \leq 10^{-10}$ atm.

**Conclusions**

Under operational conditions of a fuel cell cathode, the cubic MnO$_2$ surface with adsorbed O atoms (with a considered coverage of 12 %) is the most stable (in particular, more stable than (110) surfaces).

We have demonstrated that LaMnO$_3$ (001) surface could be catalytically more active than isostructural SrTiO$_3$ since it permits dissociative O$_2$ desorption on surface Mn ions without assistance of surface defects. The penetration of adsorbed O atoms into the first plane of cathode occurs very likely when it meets highly mobile surface O vacancies. The vacancy migration energy (0.67 eV) along the surface is smaller than in the bulk (0.95 eV) and enables fast surface diffusion of oxygen in the first cathode layer (rather than in the adsorbed state with the activation energy exceeding 1.6 eV). We do not expect entropy effects to be large enough as to favour the diffusion in the adsorbed state over surface diffusion in the first layer; however it may seriously favour bulk diffusion at very high temperatures. Quantum MD simulations would be of great interest to clarify this point.

. **Acknowledgments**
Authors are greatly indebted to R. Merkle, D. Fuks, M. Liu, N. Kovaleva, J. Fleig, Yu. Zhukovskii and H.-U. Habermeier for many stimulating discussions.

## Notes and references

[a] Max Planck Institute for Solid State Research, Heisenbergstr., 1, D-70569, Stuttgart, Germany. Fax: 49 711689 1722; Tel: 49 711689 1721; E-mail: sofia.weiglein@fkf.mpg.de
[b] Institute for Solid State Physics, University of Latvia, Kengaraga str. 8, Riga LV-1063, Latvia. Fax: 371 7132 778; Tel: 371 7187 816; E-mail: kotomin@fkf.mpg.de